\documentclass{WileyMSP-template}

\usepackage{amssymb}
\usepackage{amsfonts} 

\usepackage[super,numbers,sort&compress]{natbib}
\usepackage{placeins}

\usepackage[colorlinks,plainpages=false,linkcolor=blue,urlcolor=blue,citecolor=blue,pdfpagemode=UseNone,pdfstartview=FitBH]{hyperref}
\usepackage{graphicx}    
\graphicspath{{./}{figure/}}
\DeclareGraphicsExtensions{.eps,.png,.pdf,.jpg}
\renewcommand{\figurename}{\textbf{Figure}}
\renewcommand{\thefigure}{\textbf{\arabic{figure}}}
\DeclareUnicodeCharacter{223C}{~}

\def\EF{$E_\mathrm{F}$}

\begin{document}

\pagestyle{fancy}
\rhead{\includegraphics[width=2.5cm]{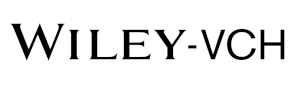}}

\title{UOTe: Kondo-interacting topological antiferromagnet in a van der Waals lattice}

\maketitle


\author{Christopher Broyles†}
\author{Sougata Mardanya†}
\author{Mengke Liu†}
\author{Junyeong Ahn}
\author{Thao Dinh}
\author{Gadeer Alqasseri}
\author{Jalen Garner}
\author{Zackary Rehfuss}
\author{Ken Guo}
\author{Jiahui Zhu}
\author{David Martinez}
\author{Du Li}
\author{Yiqing Hao}
\author{Huibo Cao}
\author{Matt Boswell}
\author{Weiwei Xie}
\author{Jeremy G. Philbrick}
\author{Tai Kong}
\author{Li Yang}
\author{Ashvin Vishwanath}
\author{Philip Kim}
\author{Su-Yang Xu*}
\author{Jennifer E. Hoffman*}
\author{Jonathan D. Denlinger*}
\author{Sugata Chowdhury*}
\author{Sheng Ran*}



\begin{affiliations}
C. Broyles, Z. Rehfuss, K. Guo, J. Zhu, D. Martinez, D. Li, L. Yang, S. Ran\\
Department of Physics, Washington University in St. Louis, St. Louis, MO 63130, USA\\
Email Address: rans@wustl.edu\\

S. Mardanya, G. Alqasseri, J. Garner, S. Chowdhury \\
Department of Physics and Astronomy, Howard University, Washington, DC 20059, USA\\
Email Address: sugata.chowdhury@howard.edu\\

M. Liu, J. Ahn, A. Vishwanath, P. Kim, J. E. Hoffman\\
Department of Physics, Harvard University, Cambridge, MA 02138, USA\\
Email Address: jhoffman@g.harvard.edu

T. Dinh, S. Xu \\
Department of Chemistry and Chemical Biology, Harvard University, Cambridge, MA 02138, USA \\
Email Address: suyangxu@fas.harvard.edu

Y. Hao, H. Cao \\
Neutron Scattering Division, Oak Ridge National Laboratory, Oak Ridge, TN 37831, USA \\

M. Boswell, W. Xie \\
Department of Chemistry, Michigan State University, East Lansing, MI 48824, USA \\

J. G. Philbrick, T. Kong \\
Department of Physics, University of Arizona, Tucson, AZ 85721, USA \\

J. D. Delinger \\
Advanced Light Source, Lawrence Berkeley National Laboratory, Berkeley, CA 94720, USA \\
Email Address: jddenlinger@lbl.gov

\end{affiliations}


\keywords{topological, antiferromagnet, van der Waals}

\begin{abstract}

Since the initial discovery of two-dimensional van der Waals (vdW) materials, significant effort has been made to incorporate the three properties of magnetism, band structure topology, and strong electron correlations — to leverage emergent quantum phenomena and expand their potential applications.  However, the discovery of a single vdW material that intrinsically hosts all three ingredients has remained an outstanding challenge. Here we report the discovery of a Kondo-interacting topological antiferromagnet in the vdW 5$f$ electron system UOTe. It has a high antiferromagnetic (AFM) transition temperature of 150~K, with a unique AFM configuration that breaks the combined parity and time reversal (\(PT\)) symmetry in an even number of layers while maintaining zero net magnetic moment. Our angle-resolved photoemission spectroscopy (ARPES) measurements reveal Dirac bands near the Fermi level, which combined with our theoretical calculations demonstrate UOTe as an AFM Dirac semimetal. Within the AFM order, we observed the presence of the Kondo interaction,
as evidenced by the emergence of a 5$f$ flat band near the Fermi level below 100 K and hybridization between the Kondo band and the Dirac band. Our density functional theory calculations in its bilayer form predict UOTe as a rare example of a fully-compensated AFM Chern insulator.

\end{abstract}


\newpage
\section{Introduction}

The current study of the topological properties of vdW materials is mainly based on a non-interacting, single-ion picture. The introduction of magnetism and strong electron correlations is expected to produce novel topological properties and other exotic phases by breaking symmetries, inducing new types of order, and creating intricate interplay between charge, spin, and orbital degrees of freedom. In vdW materials, this is typically achieved by twisting adjacent layers~\cite{sboychakov2015electronic, kerelsky2019maximized, tarnopolsky2019origin, andrei2020graphene, tran2020moire, naik2018ultraflatbands, zhang2020flat, lisi2021observation} or by utilizing specific lattice geometries~\cite{lin1994, Guo2009, jiang2023kagome, zhang2021recent, yin2022topological}. A recent milestone along this direction is the discovery of the fractional quantum anomalous Hall (FQAH) effect in ferromagnetic heterostructures, which enables the study of fractional charge and non-Abelian anyons~\cite{Park2023,Lu2024}. A wealth of exotic phenomena, including topological Kondo insulators, topological Mott insulators, chiral quantum spin liquids, and both integer and fractional quantum anomalous Hall effects with AFM order, is yet to be realized in vdW materials. Achieving these new quantum phases likely requires developing new vdW materials and novel methods to intrinsically integrate strong correlation, magnetism, and topological characteristics. 

A promising route to naturally incorporate both magnetism and strong correlation arises from the competition between the Kondo effect and magnetic order, where localized magnetic moments interact with conduction electrons to form a many-body entangled state or align via exchange interactions \cite{jullien1977, read1984, tsunetsugu1997, coleman2015heavy}. The recently identified vdW Kondo AFM material CeSiI~\cite{Posey2024} marks the transition of Kondo physics from traditional 3D to 2D. However, it does not exhibit topological properties. To search for these three attributes in a single vdW material, we focus on heavy 5$f$ elements. First, heavy elements have strong spin-orbit coupling (SOC), which can induce topological phases. Furthermore, the dual nature of 5$f$ electrons, which can exhibit either localization or itinerancy, is more promising than the typically-localized nature of 4$f$ electrons.  In particular, uranium compounds often possess multiple 5$f$ electrons, with some contributing to magnetic ordering while others engage in Kondo screening~\cite{Giannakis2019,Chen2019}.  Indeed, Kondo coherence has been observed to emerge below the magnetic ordering temperature in a few U-based compounds~\cite{Giannakis2019,Chen2019,Siddiquee2023}, making U-based systems an ideal platform for probing the rich physics at the convergence of topology, magnetism, and strong correlation.



Here we identify the vdW material UOTe as a Kondo-interacting topological antiferromagnet. It features a unique antiferromagnetic structure that naturally breaks the combined parity and time reversal ($PT$) symmetry in an even number of layers while maintaining zero net magnetization, a simple Fermi surface with Dirac-like crossings close to the Fermi level, and Kondo interactions inside the AFM state. Moreover, its vdW nature offers potential for the investigation and manipulation of a variety of exotic quantum phases in the 2D limit, such as magnetic and electric field-tunable topological phases and an intrinsic fully-compensated AFM Chern insulator in the bilayer form.

\section{Magnetic structure of the AFM state} \label{AFM}

\begin{figure*}[tbh]
\centering
\includegraphics[width=17cm]{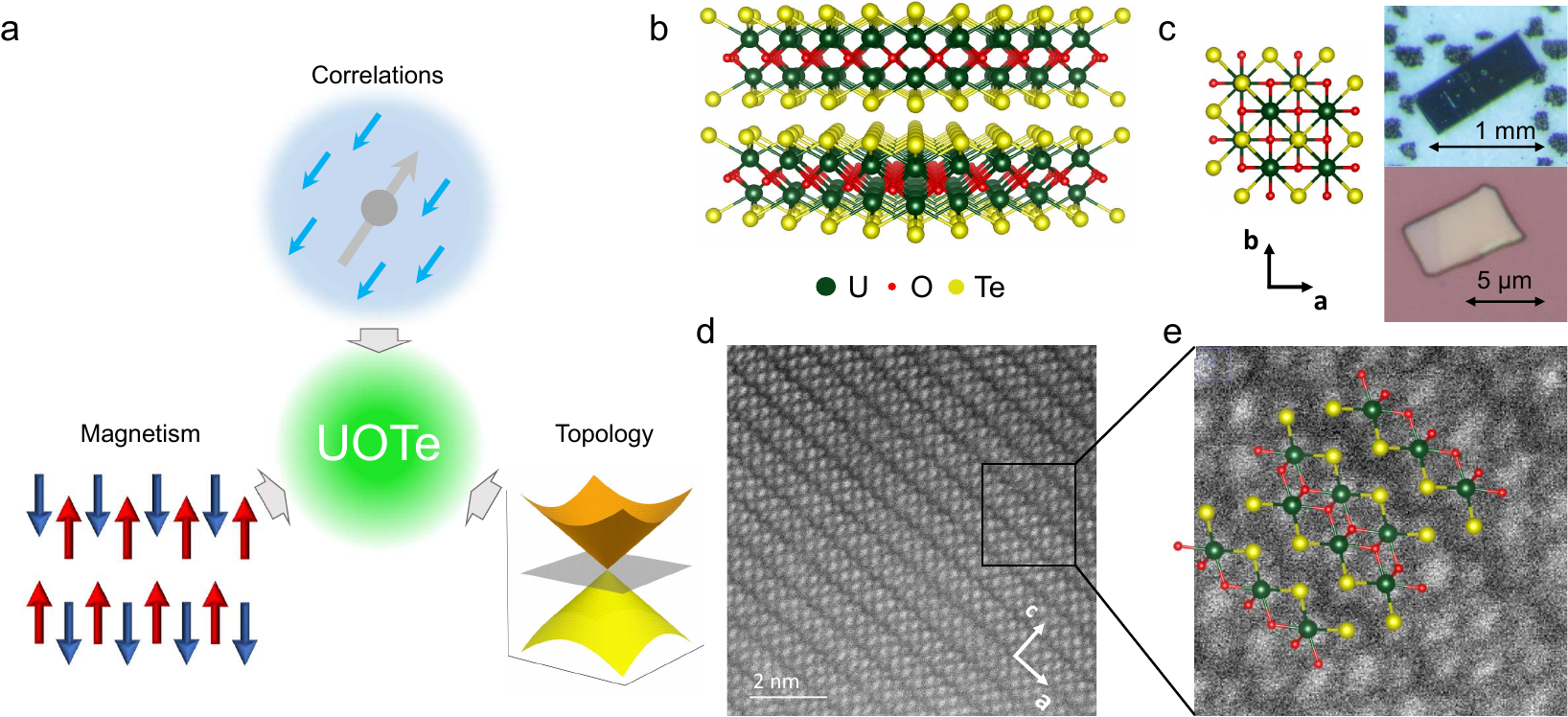}
\caption{
(a) UOTe integrates electron correlations, band structure topology, and magnetism. (b) Cross-sectional view of the UOTe crystal structure. (c) Top view of the UOTe crystal structure and bulk and flake samples of single crystal UOTe. (d) Atomically resolved cross-sectional high angle annular dark field STEM image of UOTe. (e) Magnified STEM image with atoms overlaid.
}
\label{fig1}
\end{figure*}

\begin{figure*}[tbh]
\includegraphics[width=\textwidth]{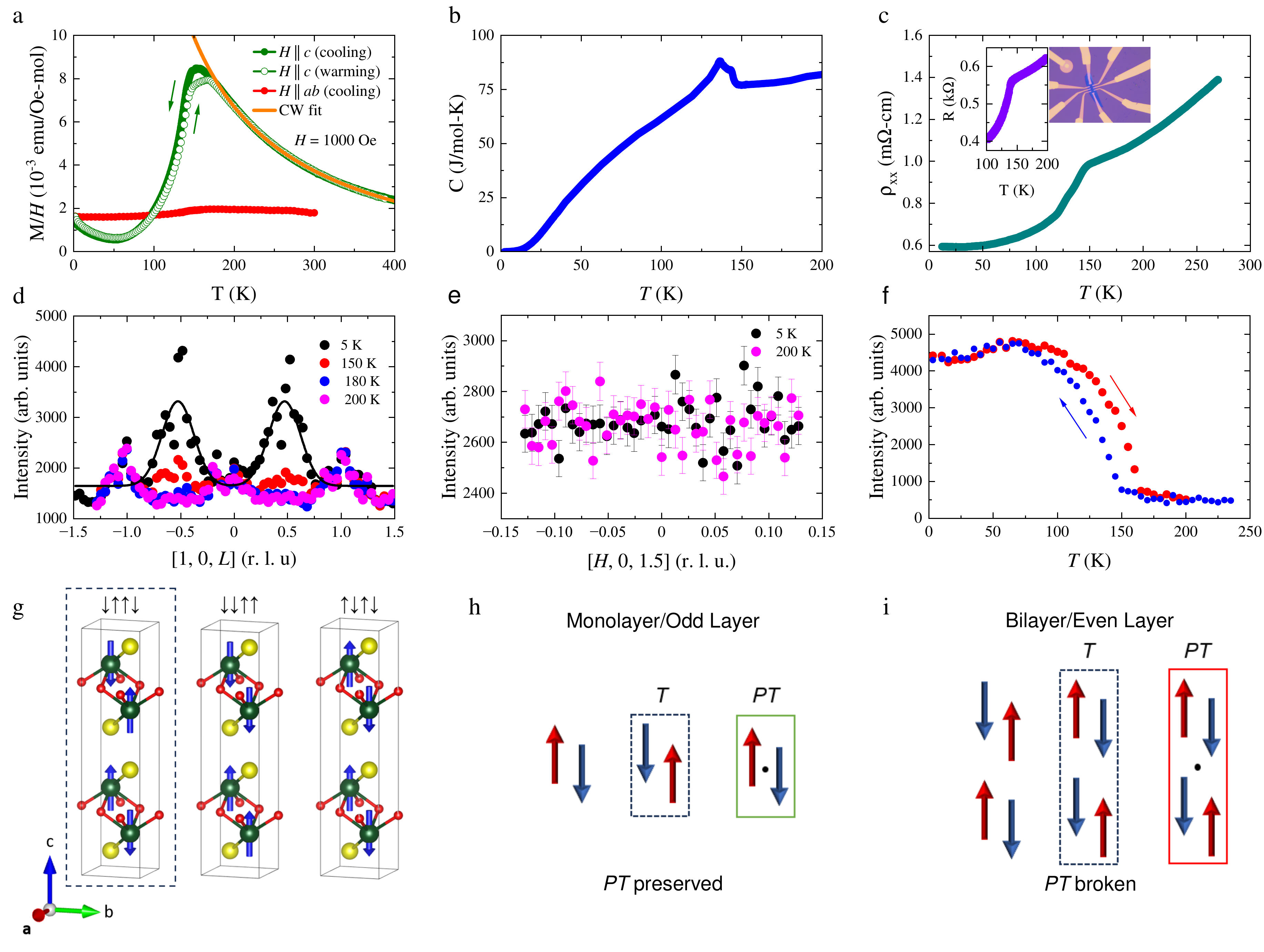}

\caption{
(a) Temperature dependence of magnetization with field of 1000 Oe field along $c$ axis and in $ab$ plane. Magnetization hysteresis is observed for the $c$ axis field upon warming and cooling. Magnetization in $ab$ plane is measured upon cooling. (b) Temperature dependence of specific heat. (c) Temperature dependence of in-plane resistivity $\rho _{xx}$ of bulk sample. Inset: in plane resistance $R_{xx}$ of a 25 nm flake.
(d) Neutron diffraction rocking curve near $\textbf{Q}$ = (1, 0, \(L\)) measured at $T$ = 5~K to 200~K. Magnetic peaks are found at \(L\) = ±0.5, but not at \(L\) = 0. Solid lines show fits to Gaussian peaks. (e) Neutron scattering intensity near $\textbf{Q}$ = (\(H\), 0, 1.5) measured at $T$ = 5~K to 200~K. No magnetic peak is found at
 $\textbf{Q}$ = (0, 0, 1.5). (f) Temperature-dependence of the $\textbf{Q}$ = (1, 0, 0.5) magnetic peak. 
The temperature points were measured in 5 K steps. Each temperature point was integrated for 1 minute after waiting 1 minute to reach equilibrium. (g) The symmetry-allowed magnetic structures with $\textbf{k}$ = (0, 0, 0.5): $\downarrow \uparrow \uparrow \downarrow$ and $\downarrow \downarrow \uparrow \uparrow$. The $\downarrow \uparrow \uparrow \downarrow$ structure agrees with the measured neutron data at 5~K. 
The $\uparrow \downarrow \uparrow \downarrow$ magnetic structure, with the magnetic space group P4/n$'$m$’$m$’$, is shown for comparison, which belongs to a magnetic symmetry with $\textbf{Q}$ = (0, 0, 0) and is not consistent with the measured data on UOTe. (h) The operation of $PT$ symmetry for a monolayer of UOTe. (i) The $PT$ operation for a bilayer configuration, breaking $PT$ symmetry. 
}
\label{fig2}
\end{figure*}

UOTe was first studied in the 1960s~\cite{Trzebiatowski1961,Stalinski1963,Murasik1965,Murasik1969}, and single crystals have been synthesized since the 1990s~\cite{Shlyk1995}. It crystallizes in the space group P4/nmm (No. 129), with the lattice parameter $a$ = $b$ = 3.952(2) Å and $c$ = 8.083(7)~Å. Its structure has a stacking sequence of quintuple layers of Te-U-O-U-Te (\textbf{Figure~\ref{fig1}b,c}). Each quintuple layer is bound to its neighboring layers via vdW forces between Te atoms, allowing for cleavage along these planes. This structural arrangement is depicted in our cross-sectional scanning transmission electron microscopy (STEM) image (Figure~\ref{fig1}d,e). The ground state of UOTe has previously been determined to be antiferromagnetic, but the exact magnetic structure remains unclear~\cite{Murasik1969,Shlyk1995}. At high temperatures, the magnetization exhibits Curie-Weiss behavior (\textbf{Figure~\ref{fig2}a}), with a Curie-Weiss temperature of -30~K and an effective magnetic moment of 3.2 $\mu_B$, slightly smaller
than that of U$^{3+}$ and U$^{4+}$. A paramagnetic to an antiferromagnetic phase transition is shown in the temperature-dependent magnetization, resistivity and specific heat measurement (\textbf{Figure~\ref{fig2}a-c}), with a transition temperature of 150~K. This transition temperature remained almost the same down to \textasciitilde 25nm of UOTe~(Figure~\ref{fig2}c inset).

To understand the magnetic order in UOTe, we employed single-crystal neutron diffraction. The crystal structure above 200~K agrees with the space group P4/nmm. Magnetic peaks appear upon cooling, with a propagation vector of \(\textbf{k}= (0, 0, 1/2)\). Figure~\ref{fig2}d displays the \(L\)-scan at (1, 0, 0) for selected temperatures ranging from 5 K to 200 K. Both the (1, 0, 1/2) and (1, 0, -1/2) peak intensities increased below 180 K. We tracked the (1, 0, 1/2) peak during warming and cooling cycles to reveal a hysteresis, which is also observed in our bulk magnetization data~(Figure~\ref{fig2}f). We then performed  \(H\)-scan along (\(H\), 0, 1.5) but we did not observe magnetic peaks at the scattering vector (0, 0, 1.5), indicating that the ordered moments align parallel to the \(c\)-axis, since neutrons detect only the magnetic moment perpendicular to the scattering vector~(Figure~\ref{fig2}e). Consequently, only two possible magnetic structures are viable, both with the magnetic space group \(P_c4/ncc\) but with different origin shifts, as depicted in Figure~\ref{fig2}g. In the down-up-up-down ($\downarrow \uparrow \uparrow \downarrow$) ordering, the nearest U-moments exhibit antiferromagnetic coupling, whereas in the down-down-up-up ($\downarrow \downarrow \uparrow \uparrow$) ordering, they display ferromagnetic coupling. We used FullProf~\cite{RodriguezCarvajal1993} refinement to confirm that only the $\downarrow \uparrow \uparrow \downarrow$ structure aligns with the measured data at 5 K. Alternatively, interlayer antiferromagnetic coupling could potentially result in another antiferromagnetic order with the magnetic space group P4/n$’$m$’$m$’$, namely up-down-up-down ($ \uparrow \downarrow \uparrow \downarrow$). While this order also exhibits antiferromagnetic coupling between the next nearest U-moments, it corresponds to a propagation vector of \(\textbf{k}= (0, 0, 0)\). This magnetic structure has the strongest magnetic peak at $\textbf{Q}$ = (1, 0, 0), which was not observed for UOTe. Therefore, we conclude that the magnetic ground state of UOTe possesses the $\downarrow \uparrow \uparrow \downarrow$ structure, which is classified by the magnetic space group P$_c$4/ncc (130.432), with inversion $\mathbb{I}$, horizontal mirror $\mathbb{M}_{001}$, and rotation with translation $\{\mathbb{C}_{4z}|\frac{1}{2} 0 0\}$. 

This $\downarrow \uparrow \uparrow \downarrow$ magnetic structure leads to intriguing consequences. The spin configuration remains invariant under the combined parity and time-reversal ($PT$) symmetry operation, in both the bulk material and odd quintuple layers of Te-U-O-U-Te, as shown in Figure~\ref{fig2}h. The inversion center is between the up and down spins. In contrast, in the even number of quintuple layers, the $PT$ symmetry is broken because the system is now even under parity $P$ but odd under time-reversal $T$~(Figure~\ref{fig2}i). This symmetry breaking allows for a non-zero Chern number despite the overall zero net magnetic moment, holding significant promise towards achieving AFM QAH effect. Proposals exist to achieve AFM QAH by engineering magnetic topological insulators such as MnBi$_2$Te$_4$ into such a magnetic structure~\cite{Liang2024}; however we will demonstrate that such exotic quantum phenomena can emerge naturally in UOTe, given its inherent magnetic structure and topological properties.

\section{Antiferromagnetic Dirac semimetal}

\begin{figure*}[tbh]
\centering
\includegraphics[width=10.5cm]{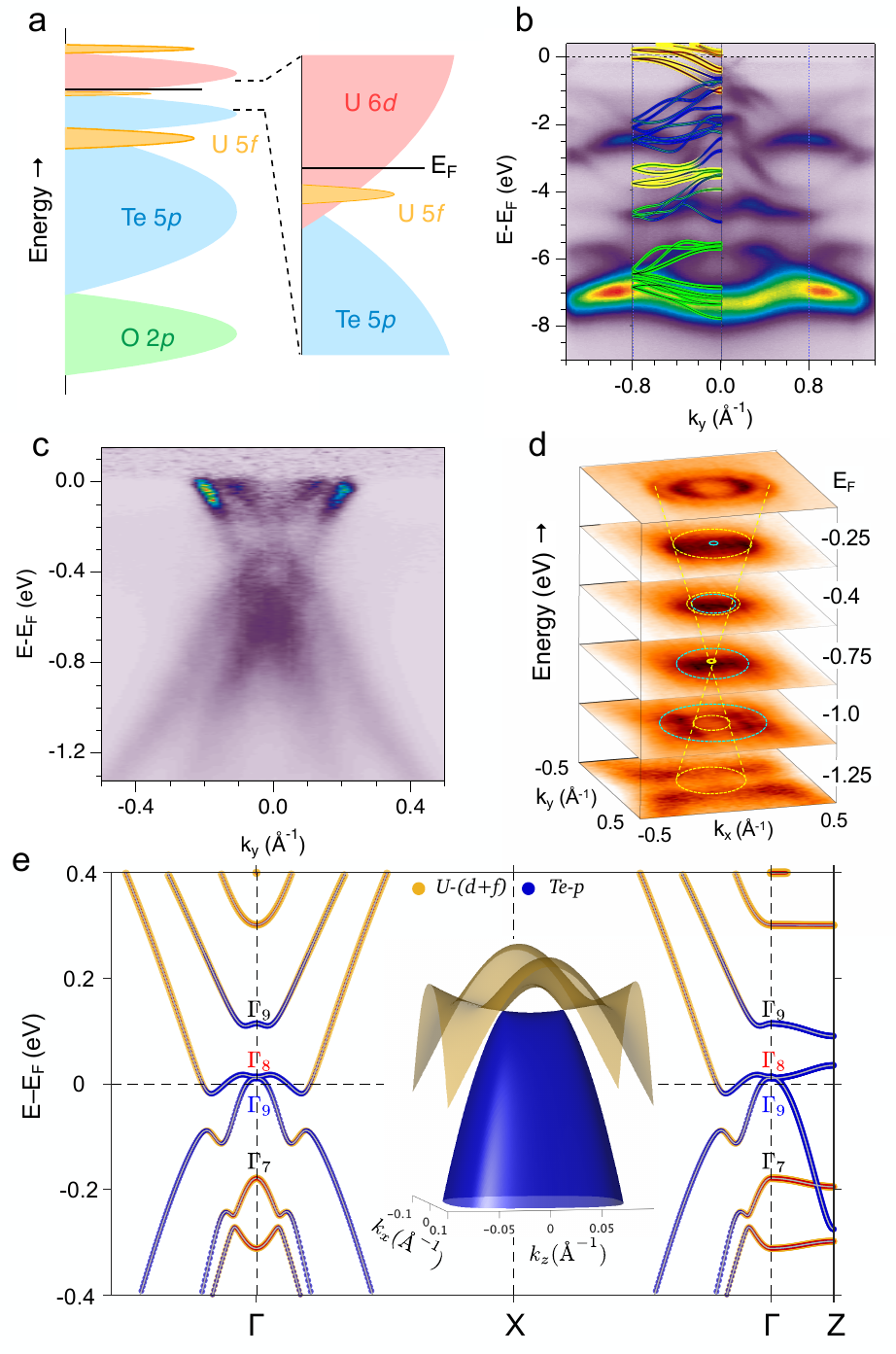}   
\caption{
(a) Simplified schematic of the electronic orbitals with zoom in the Fermi-level.
(b) ARPES measurement of wide valence band dispersion using 90 eV photon energy with LH + LV polarizations.  Comparison is made to a DFT+$U$ ($U$=4 eV) calculation for bulk UOTe with SOC. The agreement is achieved only after shifting the DFT bands down by ∼ 0.5 eV. Band character color-coding: Te-$p$ (blue), U-$f$ (yellow), U-$d$ (red), and O-$p$ (green).
(c) Energy normalized normal emission 90 eV valence  band image for LH + LV polarizations. 
(d)  
ARPES map measured at 35 eV for Fermi-edge and various binding energies.
(e) Electronic structure of bulk UOTe with SOC using $U$=4 eV along the high symmetry path of the BZ. The colormap represents the orbital character of U-$(d+f)$ and Te-$p$ at each point. A Dirac crossing appears along $\Gamma-Z$ direction. The Dirac crossing is rotation symmetry protected as the band inversion happens between the $\Gamma_8$ and $\Gamma_9$ bands which have opposite $C_{4z}$ eigenvalues.
}
\label{fig3}
\end{figure*}

We now study the band structure and topology of UOTe. As is well-established, nontrivial topology typically requires an energy inversion between the conduction and valence bands. Depending on whether the inverted bands are gapped everywhere in $k$-space or remain gapless at certain $k$-points, a topologically nontrivial insulator or semimetal is formed. Before presenting our ARPES data, we provide an overview of the UOTe electronic structure. As shown in \textbf{Figure\ \ref{fig3}a}, far from the Fermi level (\EF), the O 2$p$ valence band is present, as well as the U 5$f$ flat bands. Near the Fermi level, the valence band and conduction band correspond to Te 5$p$ and U 6$d$, respectively. Crucially, there is a band inversion between the Te 5$p$ and U 6$d$ bands, which forms the foundation of the nontrivial topology. Additionally, there are weak flat features near \EF, shown in the zoomed-in view of Figure\ \ref{fig3}a, which are attributed to Kondo-resonance effects that will be discussed in the next section.
Figure\ \ref{fig3}b shows the ARPES-measured band structure over a wide energy window with comparison to DFT+$U$ ($U$ = 4.0~eV) calculated bands. The agreement is achieved only after shifting the DFT bands down by $\sim$ 0.5 eV, indicating a band-filling of the electronic structure in the sample. This is most straightforwardly rationalized by the presence of off-stoichiometry and impurities in the as-grown material as characterized by SEM-EDX (see Supplementary Fig. S1 and Table S1).

At higher binding energies, our data reveals many valence bands, which can be nicely reproduced by our DFT+$U$ calculations. However, near the Fermi level, we observe small Fermi pockets near the $\Gamma$ point.

To investigate the details of the electronic band structure close to \EF, we present an ARPES dispersion map in Figure\ \ref{fig3}c. Remarkably, we observe a clear Dirac cone that connects the conduction and valence bands, with no observable gap. Within the Dirac cone, there is a secondary inner band. The presence of a gapless Dirac cone can be explained by two topological states, an antiferromagnetic topological insulator or an antiferromagnetic Dirac semimetal. In the former case, the Dirac cone will be a two-dimensional surface state, whereas in the latter case, the Dirac cone will be a three-dimensional bulk band. In order to distinguish these two cases, we perform photon energy $k_z$ dependent studies. As shown in \textbf{Figure~S1}, 
we observe clear $k_z$ dependence, therefore, our ARPES data demonstrates UOTe as an antiferromagnetic Dirac semimetal.




Our DFT+$U$ calculations further illuminate the topological nature of UOTe. At $U$=4 eV (value fixed by comparison with ARPES data), our DFT+$U$ predicts that UOTe is an antiferromagnetic Dirac semimetal, where the bulk Dirac fermion is protected by the $C_4$ symmetry. The bands near the (stoichiometric) Fermi energy at the $\Gamma$ point are characterized by the Irreducible Representations (IRR) $\Gamma_9$ and $\Gamma_8$, respectively. These bands only differ in $C_4$ rotation eigenvalues and remain gapped in the Brillouin zone (BZ) except along the $\Gamma$-$Z$ direction. Along the $\Gamma$-$Z$ direction, these two doubly degenerate bands cross each other to form a tilted Dirac crossing, with the gapless point protected by the $C_4$ rotation symmetry. The Dirac points are located 10.6 meV above the (stoichiometric) Fermi energy at $(0, 0, \pm0.015 )$ \AA$^{-1}$. We plotted the three-dimensional dispersion over the $k_x$-$k_z$ plane, highlighting the tilted nature of these Dirac crossings.

Although our DFT+$U$ calculations successfully capture the qualitative features of the band structure topology, some fine details exhibit discrepancies when compared to ARPES measurements. For instance, the calculated bands near the Fermi level show less dispersion, leading to a more quadratic than linear dispersion around the Dirac crossing. This deviation is typical for strongly correlated systems, where DFT+$U$ often has limitations. To achieve a more precise understanding of the topology of UOTe, more advanced methods such as dynamical mean-field theory are required.



\section{Kondo many-body effect} \label{ARPES2}

\begin{figure*}[ht]
\begin{center}
\includegraphics[width=17cm]{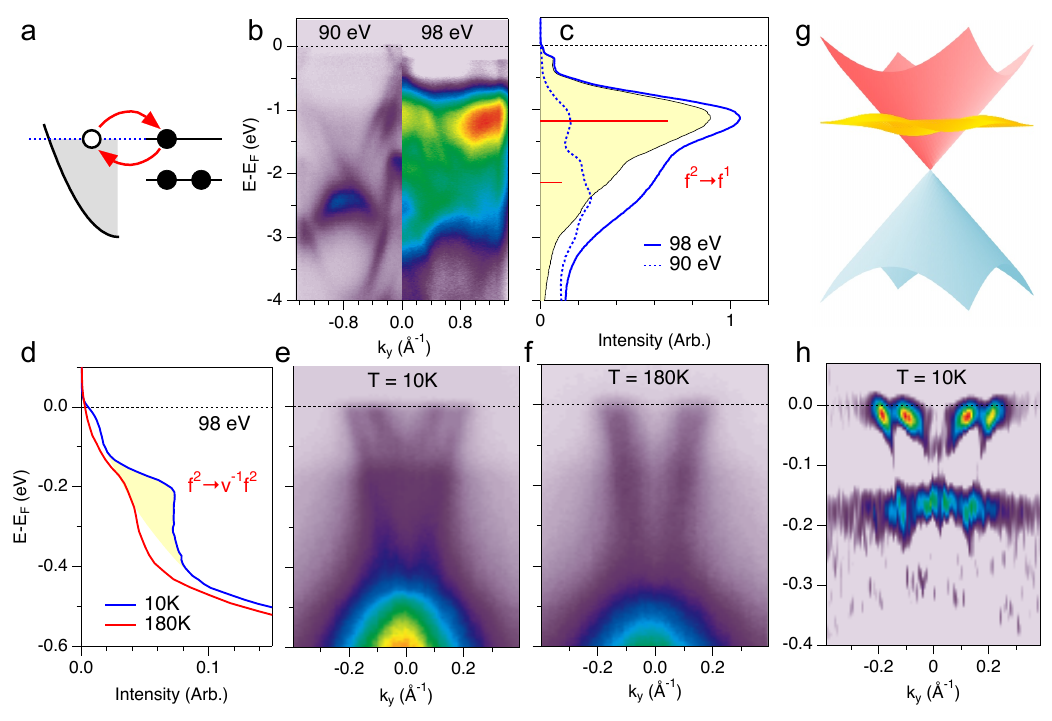} 
\caption{
(a) Simplified schematic of Kondo hybridization hopping between conduction band and U 5$f$ electrons with a fluctuating valence between $f^2$ and $f^3$. 
(b) Comparison of off-resonance 90 eV and on-resonance 98 eV valence band images, both taken at 10K, with angle-integrated spectra in (c) highlight the enhanced U 5$f$ spectral weight in the yellow-shaded difference spectrum. 
Horizontal red lines in (c) are predicted relative energies and intensities of two spin-orbit split $f^2$$\rightarrow$$f^1$ final states.
(d) Near-\EF\ on-resonance angle-integrated spectra at 10 K and 180 K showing an additional U 5$f$ peak at -0.2 eV (yellow shaded region) with weakened intensity at high temperature.
(e,f) Normal emission valence band spectra at 35 eV using LV-polarization at (e) 10 K and (f) 180 K, showing temperature dependence of both the Dirac-like electron bands and the flat -0.2 eV U 5$f$ peak. 
(g) Schematic of the $f$-electron flat band coexisting with the Dirac-like conduction band. 
(h) Second derivative of the 35 eV spectra at 10K along the energy direction, highlighting the flat $f$-band and its interior enhancement due the hybridization with Dirac band.
}
\label{fig4}
\end{center}
\end{figure*}

The Kondo resonance is a many-body effect resulting from the hybridization between localized magnetic moments and itinerant conduction electrons. In uranium compounds, 5$f$ electrons have a dual nature: being both itinerant and localized. As shown in \textbf{Figure~\ref{fig4}a}, the band structure includes itinerant bands crossing the Fermi energy, localized 5$f$ states well below the Fermi energy, and hybridizable 5$f$ states near the Fermi energy. The hybridization between conduction electrons and the 5$f$ states near the Fermi energy leads to Kondo resonances, appearing as sharp peaks in the density of states near the Fermi energy. Kondo resonances are beyond the single-particle band structure and serve as a distinct signature of the many-body interactions in the system.

Before presenting evidence of the Kondo many-body effect in UOTe, we first show the distribution of the U 5$f$ spectral weight. Figure~\ref{fig4}b shows a comparison of valence band dispersion using 90~eV photoexcitation with that using 98~eV photon which couples resonantly to the U 5$d$$\rightarrow$5$f$ transition, thus strongly enhancing the photoemission cross section of the $f$ states. Figure~\ref{fig4}c shows the angle-integrated off-resonance 90 eV and on-resonance 98 eV spectra, along with a difference spectrum highlighted with yellow shading. Here we observe the U 5$f$ spectral weight composed of a primary broad peak at -1 eV and a similarly broad shoulder at about -2 eV. In addition, a weaker peak at -$0.2$ eV is observed with 98 eV resonant enhancement at low temperatures, indicating its origin from U 5$f$ electrons, as highlighted in the zoomed-in energy scale plot of Figure\ \ref{fig4}d.

While the strong $f$-peaks at -1~eV and its shoulder around -2~eV correspond to the very localized states (\textbf{Figure~S2}),
the weaker peak at -$0.2$~eV, not previously reported in localized 5$f$ systems like UO$_2$~\cite{Baer1980,Roy2008} or UPd$_3$ \cite{Allen2000, Kawasaki2013}, highlights the Kondo-interacting many-body effect. Temperature dependence can be used to distinguish these two scenarios, as localized $f$ bands typically do not exhibit temperature dependence, while Kondo-entangled bands do, and will only emerge below the Kondo coherence temperature. As shown in Figure\ \ref{fig4}d, the -0.2 eV peak present at 10~K is greatly suppressed at 180~K, suggesting the Kondo-interacting nature. The flat band at -0.2 eV and its $T$-dependence is also observed in 35~eV spectra in Figure\ \ref{fig4}e and f. Additional temperature-dependent peak intensity evolution shows this flat band emerges below 100~K 
(\textbf{Figure~S3}).
Furthermore, the hybridization between the $f$ states and the conduction bands is evidenced in the second derivative image of 35~eV at 10~K in Figure\ \ref{fig4}h which highlights the interior enhancement of the flat band due the hybridization. Even clearer evidence of hybridization is seen in the second derivative image of 90~eV 
(\textbf{Figure~S4e}),
e.g., the inner portion of the flat band to move upward and the outer portion to move downward. This hybridization is not merely a Kondo effect but specifically involves the topological Dirac band, further demonstrating the intricate interplay between strong correlations and topological properties in UOTe. This key observation underscores the significance of UOTe as a platform for studying emergent quantum phenomena where Kondo physics intersects with topological Dirac states.


Our observation of the -0.2 eV Kondo flat band contrasts with typical Kondo lattice materials, where the Kondo resonance peak occurs near the Fermi energy. During the $f$-electron photo-emission process, the Kondo resonance is reflected as a hopping of conduction electron located at at \EF~to refill the photo-ejected $f$-electron. The result is a $5f^2$ final state, the same as the initial state configuration, but with a valence hole ($5f^2$$\rightarrow$$v^{-1}$$5f^2$). Such Kondo-screening electron hopping typically results in a strong photoemission peak at \EF ~\cite{Parks1983}. However this peak is suppressed here in UOTe, likely due to its relatively small Fermi surface resulting from its semimetal nature and the presence of magnetic order~\cite{Baer1980,Roy2008,Jang2019}. Instead, the -0.2 eV U 5$f$ peak, corresponding to a $5f^2$ final state multiplet excitation of the Kondo resonance 
(see Supporting Information Figure~S2),
similar to the Kondo spin-orbit satellite peak in Ce \cite{Reinert2001}, is more prominent. Comparable $f$-spectral behavior has been observed in the 4$f$ system of CeSb which exhibits highly localized $T$-independent 4$f$ states, and $T$-dependent Kondo ($4f^1$$\rightarrow$$v^{-1}$$4f^1$) excitation satellites, without significant 4$f$ weight at $E_F$ \cite{Jang2019}. In UOTe, this Kondo spin-orbit satellite excitation emerges well below $T_N$, suggesting the novel coexistence of Kondo interaction and strong magnetic order~\cite{Jang2019,Giannakis2019,Chen2019,Siddiquee2023,Xu2024}.

The signature of Kondo interaction is not distinctly observed in our electronic transport and heat capacity measurements. For instance, the electronic contribution to the specific heat \( C/T \) is approximately 6 mJ/mol-K\(^2\) at low temperatures. This observation aligns with the suppression of the Kondo resonance at the Fermi level and its coexistence with magnetic order. It is plausible that substituting oxygen with sulfur or selenium could decrease the charge transfer to the oxygen $p$-states, thereby enhancing the $f-p$ interaction. A promising future direction involves synthesizing USTe and USeTe in van der Waals lattice and S and Se-doped UOTe to amplify the Kondo interaction and its manifestation in transport and thermodynamic measurements. Another promising direction is to explore the ultra-thin UOTe, where the electrostatic gating may enhance the Kondo signature in transport.

\section{Conclusion} 

We have identified UOTe as the first intrinsic material exhibiting the coexistence of Kondo interactions, antiferromagnetism, and non-trivial topology. The interplay of these three ingredients gives rise to a variety of exotic phenomena. For example, the unique magnetic structure of UOTe naturally breaks $PT$ symmetry in even layers, allowing for a non-zero Chern number despite the overall zero net magnetic moment. This, combined with its topological phase, potentially leads to the formation of a fully compensated antiferromagnetic Chern insulator, which have not had material realization before. Indeed, our DFT calculations confirm that the bilayer configuration exhibits a global gap throughout the Brillouin zone and a Chern number $C$ = 1 
(\textbf{Figure~S5}).
This Chern insulating phase is computationally verified by the surface state spectrum obtained along the line connecting the corners of the surface Brillouin zone, as depicted in Figure~S5e.
At the $C$ = 1 phase, only one right-moving surface state crosses the Fermi energy, satisfying the bulk-boundary correspondence. This surface state gives rise to quantized Hall conductivity 
(Figure~S5f).
Additionally, our evidence of electron correlations from observed Kondo interactions implies UOTe holds great promise in realizing correlated topological states, including topological Kondo insulators, topological Mott insulators, chiral quantum spin liquids, and fractional states. 


\section{Experimental Section}
\threesubsection{Single crystal synthesis and characterization}\\
The growth of UOTe was performed in a two-zone furnace, using equimolar ratios of depleted uranium (sourced from New Brunswick Laboratory) and tellurium (99.9999, sourced from Alfa Aesar). The source materials  were sealed in an evacuated quartz tube, together with 1~mg/cm$^3$ iodine. The ampoule was place in a two-zone furnace and gradually heated up and held in the temperature gradient of 1030/970~$^{\circ}$C for 7 days, after which it was cooled to room temperature.
The crystal structure was determined by $x$-ray powder diffraction using a Rigaku $x$-ray diffractometer with Cu-K$_{\alpha}$ radiation. Several single crystals were picked up and examined in the Bruker D8 Quest Eco single-crystal X-ray diffractometer equipped with Photon II detector and Mo radiation ($\lambda_{K\alpha}$ = 0.71073 Å) to obtain the structural information. The crystal was measured with an exposure time of 10 s and a scanning 2$\theta$ width of 0.5° at room temperature. The data was processed in Bruker Apex III software and the structural refinement were conducted with the SHELXTL package using direct methods and refined by full-matrix least-squares on $F^2$. Electrical transport measurements were conducted using a standard lock-in technique, magnetization was measured with a vibrating sample magnetometer (VSM), and specific heat measurements were performed, all within a Quantum Design Physical Property Measurement System (PPMS).

\threesubsection{Flakes exfoliation and device fabircaiton}\\

All fabrication processes were completed in an argon-filled glovebox. The glovebox was attached to an e-beam evaporator, allowing us to deposit metal in situ. UOTe was first mechanically exfoliated onto PDMS (polydimethylsiloxane) using Scotch Magic tape and transferred onto a baked 285~nm SiO2 /Si wafer. Then a stencil mask technique was used to evaporate Cr/Au contacts on top of UOTe without exposing it to air or chemicals. A hexagonal boron nitride (hBN) flake was exfoliated onto PDMS and transferred onto UOTe to further protect UOTe flake. We wire-bonded the sample inside the glovebox and sealed the device carrier with argon before loading it into cryostat for transport measurement.


\threesubsection{DFT+U calculations}\\
The UOTe crystallizes in space group P4/nmm (number 129). The crystal structure consists of two O atoms at the Wyckoff position at 2a. The U atoms are placed at Wyckoff position 2c (z=$\pm$0.15960), forming a layer of tetrahedra with the Oxygens, respecting the inversion symmetry. These layer structures repeat along the c-axis and are separated by a weak van der Waal gap. The Te atoms are placed in between these UO layers at Wyckoff position 2c (z=$\mp$0.35280), which makes Te a natural termination. 
To get insight into the electronic structure of the UOTe, we have employed the first-principle calculations within the framework of the density functional theory (DFT) \cite{hohenberg1964inhomogeneous} using the Vienna {\it ab-initio} simulation package (VASP) \cite{furthmuller1996dimer, kresse1996efficient, kresse1999ultrasoft}. The ground state electronic structure is obtained with the projector augmented-wave pseudopotential, while the electron exchange-correlation effects are considered by the Local density approximation (LDA). The energy cut-off of 500 eV was used for the plane-wave basis set, and the Brillouin zone (BZ) integration was performed with a 13$\times$13$\times$6 $\Gamma$-centered $k-$mesh \cite{monkhorst1976special}. The total energy tolerance criteria are set to $10^{-8}$ eV to satisfy self-consistency. We used the experimental structure parameters while the ionic positions were optimized until the residual forces on each ion were less than 10$^{-2}$ eV/\AA $~$ and the stress tensors became negligible. The weak interlayer van der Wall interaction is treated by the DFT-D3 correction of Grimme with zero-damping function \cite{grimme2010consistent}. The strong correlation effect of the valance U-f orbitals is corrected by considering an effective onsite Hubbard potential ($U_{\rm eff}$) \cite{hubbardU, Anisimov_1997}. 

\threesubsection{ARPES}\\ 
Angle-resolved photoelectron spectroscopy (ARPES) measurements were performed at the MERLIN beamline 4.0.3 of the Advanced Light Source (ALS) in the energy range of 30-124 eV employing both linear horizontal (LH) and linear vertical (LV) polarizations from an elliptically polarized undulator. 
A Scienta R8000 electron spectrometer was used in combination with a six-axis helium cryostat goniometer in the temperature range of 10-200K with a total energy resolution of $\geq$15 meV and base pressure of $<$5$\times$10$^{-11}$ Torr. 
Single crystal platelets of UOTe were razorblade cut to $\sim$2x2 mm lateral size, top-post cleaved in UHV, and the flattest uniform regions probed with an $\sim$50 $\mu$m beam spot size.
The Te 4d core-level exhibits a two-component structure whose low binding energy peak 
(\textbf{Figure~S6})
is consistent with cleavage between Te-termination quintuple layers.

\threesubsection{Neutron diffraction}\\ 
Neutron diffraction measurements were performed at HB3A DEMAND at High Flux Isotope Reactor at Oak Ridge National Lab \cite{Chakoumakos2011}. The experiment used Si (220) monochromator with the wavelength of 1.542 Å \cite{cao2018demand}. The experiment used coaligned single crystals on a Si substrate. The coaligned sample has total mass of ~0.1g and mosaicity of ±5°. The sample was mounted on the four-circle goniometer and cooled down to 5 K using a helium closed cycle refrigerator. The data reduction used ReTIA \cite{Hao2023}. The symmetry analysis used Bilbao Crystallography Server \cite{Aroyo2006}. The structure refinement used Fullprof \cite{RodriguezCarvajal1993}.

\medskip
\textbf{Supporting Information} \par 
Supporting Information is available from the Wiley Online Library or from the author.

\medskip
\textbf{Acknowledgements} \par 
We acknowledge fruitful discussions with Paul Canfield, Lei Chen, Daniel Dessau, Ni Ni, Priscila F. S. Rosa, Qimiao Si, Kai Sun, and  Jianxin Zhu. The work at Washington University is supported by the National Science Foundation (NSF) Division of Materials Research Award DMR-2236528. C. Broyles acknowledges the NRT LinQ, supported by the NSF under Grant No. 2152221. 
Sougata Mardanya and Sugata Chowdhury from Howard University, work supported by the U.S. Department of Energy (DOE), Office of Science, Basic Energy Sciences Grant No. DE-SC0022216. Gadeer Alqasseri and Jalen Garner from Howard University would like to thank the IBM-HBCU Quantum Center for financial support. This research at Howard University used the resources of Accelerate ACCESS  PHYS220127 and PHYS2100073. 
M. Liu acknowledges the Harvard Quantum Initiative Postdoctoral Fellowship and the assistance of Austin Akey, Jules Gardener, and Timothy J. Cavanaugh from the Center for Nanoscale Systems at Harvard in conducting the EDX, SEM, and TEM measurement. 
J. Ahn was supported by the Center for Advancement of Topological Semimetals, an Energy Frontier Research Center funded by the U.S. Department of Energy Office of Science, Office of Basic Energy Sciences, through Ames Laboratory under contract No. DE-AC02-07CH11358.
The single crystal X-ray structure determination was supported by the U.S. DOE Basic Energy Sciences via the grant DE-SC0023648. D.L. and L.Y. are supported by the Air Force Office of Scientific Research (AFOSR) Grant No. FA9550-20-1-0255. The simulation used Purdue Anvil CPU at Purdue University through allocation DMR100005 from the Advanced Cyberinfrastructure Coordination Ecosystem: Services $\&$ Support (ACCESS) program. Research at the University of Arizona is supported by the NSF under Award No. DMR-2338229. Photoemission used resources of the Advanced Light Source, which is a DOE Office of Science User Facility under contract no. DE-AC02-05CH11231. The work at Oak Ridge National Laboratory (ORNL) was supported by the U.S. DOE, Office of Science, Office of Basic Energy Sciences, Early Career Research Program Award KC0402020, under Contract DE-AC05-00OR22725. This research used resources at the High Flux Isotope Reactor, a DOE Office of Science User Facility operated by ORNL. 

C. Broyles, S. Mardanya and M. Liu have contributed equally to this work. 
\medskip

%

\bibliographystyle{MSP.bst} 



\clearpage


\setcounter{figure}{0}
\renewcommand{\figurename}{\textbf{Figure}}
\renewcommand{\thefigure}{\textbf{S\arabic{figure}}}
\renewcommand{\tablename}{\textbf{Table}}
\renewcommand{\thetable}{\textbf{S\arabic{table}}}

\FloatBarrier

\sectionmark{Photon-dependence}
\begin{figure*}[ht]
\begin{center}
\includegraphics[width=16cm]{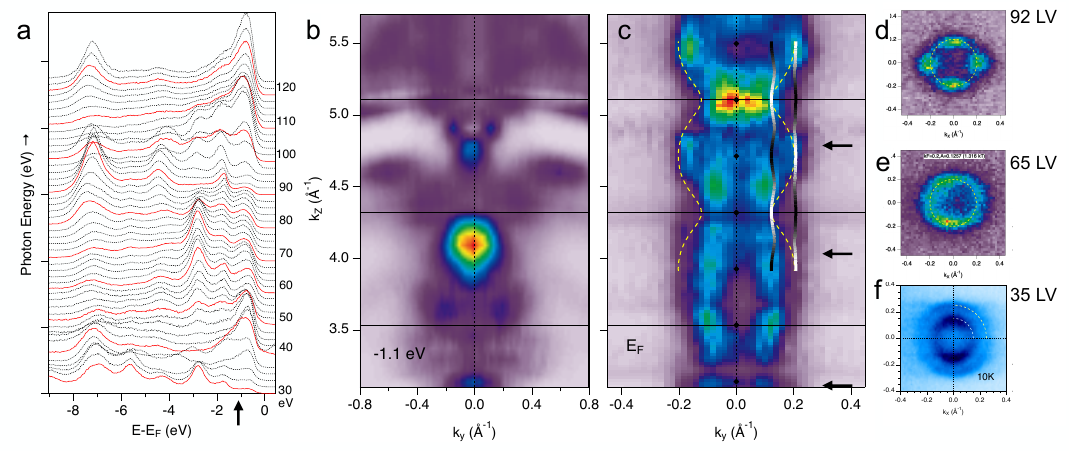}
\caption{
(a) Photon-dependence of 8 eV wide valence band spectra at normal emission from 30-124 eV using LH polarization illustrating strong variations of band intensities including the  U 5$f$ resonant enhancement above 98 eV. 
(b,c) $k_z$-dependent map using LV-polarization with energy cuts at (b) -1.1 eV illustrating strong $k_z$ dispersion, and at (c) the Fermi-edge illustrating periodic variation of the relative intensity of inner and outer electron bands. 
DFT+U constant energy Fermi surface contours for the AFM zone-folded band structure are overplotted with intensity variation according to the non-zone-folded band structure (dashed yellow line).
(d,e,f) Fermi surface maps at selected photon energies (arrows in c) illustrating the choice of 35 eV photon energy for optimized measurement of both inner and outer circular FS contours.
}
\label{ext_arpes_hv}
\end{center}
\end{figure*}

\sectionmark{f-schematics}
\begin{figure*}[ht]
\begin{center}
\includegraphics[width=10cm]{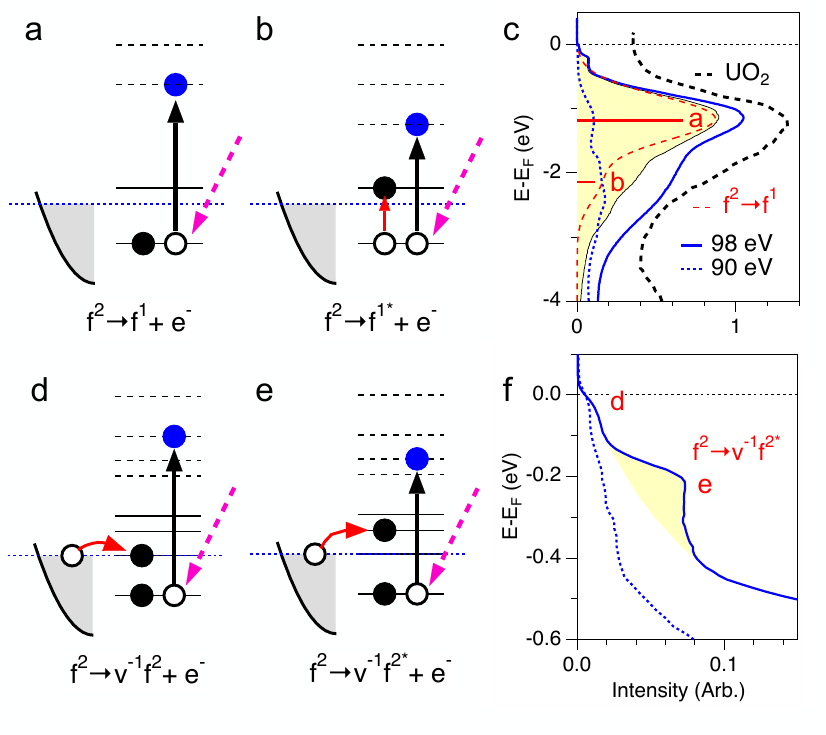}  
\caption{
Schematic energy level diagrams for one-electron removal from an $f^2$ initial state. We interpret the broad peak at -1~eV and its shoulder around -2~eV in the angle-integrated resonant photoemission difference spectrum in (c) to be the $f^1$ final-state peak(s) associated with the cost of removing one electron from the tetravalent U ion (5$f^2$$\rightarrow$5$f^1$), i.e.\ the lower Hubbard band. 
(a,b) The $f^1$ final state configurations correspond to the strong on-resonance $f$-peaks at -1.9 and -1.0 eV in the angle-integrated resonant photoemission difference spectrum in (c). An excitation into a  spin-orbit split $f^{1*}$ final state in (b) reduces the relative kinetic energy compared to (a).
Comparison is made in (c) to predictions of the relative energies ($-0.9$ eV) and intensities (6:1) of the spin-orbit split $f^1$ final state \cite{Baer1980} (red bars and red dashed curve), and also to the valence band spectrum of UO$_2$  \cite{Baer1980}. 
(d,e) The $f^2$ final state configurations result from hopping from the valence band leaving a valence hole ($v^{-1}$).
Hopping into an $f^{2*}$ excited-state multiplet in (e) is proposed to correspond to the weak flat $f$-level at -0.2 eV in (f), while the main Fermi-level Kondo resonance peak in (d) is suppressed in UOTe (see text). 
}
\label{ext_fschematics}
\end{center}
\end{figure*}

\sectionmark{AFM Band folding}
\begin{figure*}[ht]
\begin{center}
\includegraphics[width=16cm]{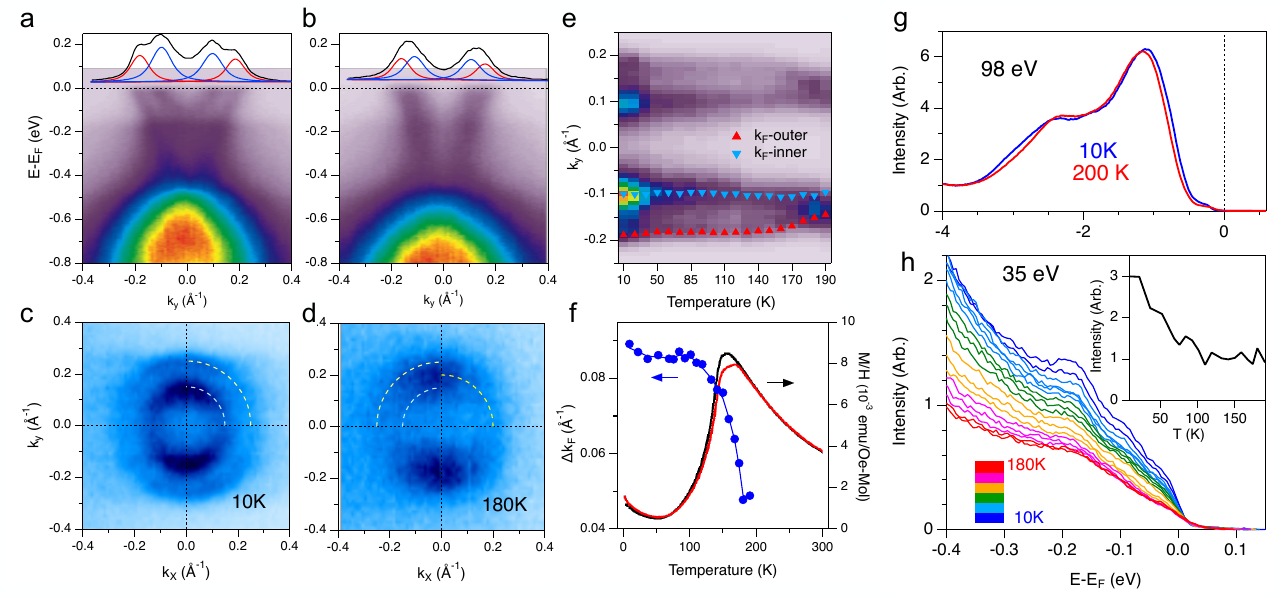}
\caption{
We investigated the temperature dependence of the normal emission spectra with warming the sample through $T_N$. 
(a,b) Normal emission valence band spectra and (c,d) Fermi surface maps at 35 eV using LV-polarization are shown for (a,c) 10K and (b,d) 180K. Fermi-energy momentum profiles are also shown with peak fitting. 
(e) Fine-step temperature dependence of the Fermi-edge momentum showing the merging of the outer and inner electron bands.
(f) Temperature-dependence of the Fermi-edge $k$-separation, 
compared to the $c$-axis magnetization profile. 
(g) Low and high temperature 98 eV spectra showing no change in the lower Hubbard band U5$f$ intensity.
(h) Temperature-dependent 35 eV normal emission spectra reveal the -0.2 eV U5$f$ peak. The intensity of the -0.2 eV peak, after a linear background subtraction, as a function of temperature is summarized in the inset which shows an increase in intensity below 100 K, indicating emergence of flat band below 100 K. }
\label{ext_arpes_tdep}
\end{center}
\end{figure*}

\sectionmark{90 eV EDC band image processing}
\begin{figure*}[ht]
\begin{center}
\includegraphics[width=14cm]{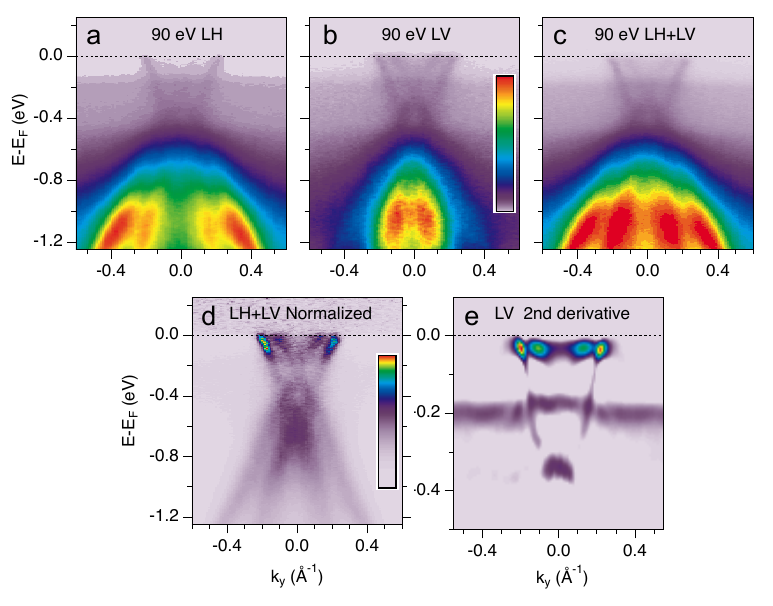}
\caption{
Normal emission valence band images measured at 90 eV with (a) LH-polarization, and (b) LV polarization. The inner electron band is selectively suppressed for LH polarization and the outer hole band is selectively suppressed for LV polarization.
(c) Sum of LH and LV polarization spectra to better represent all bands.  
(d) Intensity normalized spectra (division by a $k$-averaged spectrum) to help flatten the dynamic range, and observe the connectivity between electron and hole bands.
The flat band feature at -0.2 eV is artificially suppressed.
(e) Second derivative of the LV spectra along the energy direction to enhance heavy mass band curvatures at \EF\ and the -0.2 eV flat band including an energy shift inside the electron band region due to $f$-$d$ hybridization interactions.
}
\label{ext_arpes_90}
\end{center}
\end{figure*}


\begin{figure*}[ht]
\begin{center}
\includegraphics[width=15cm]{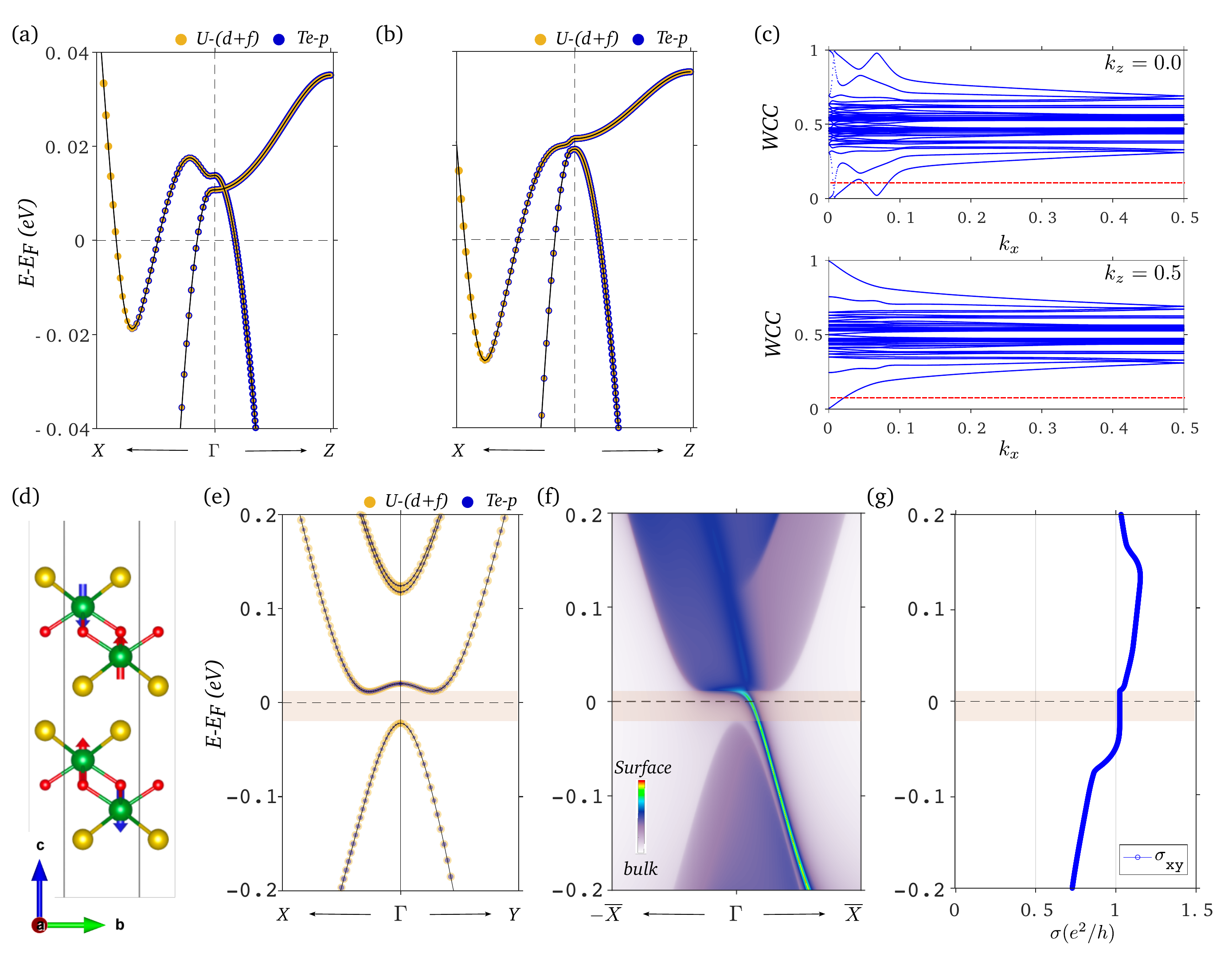}
\caption{
(a) Orbital projected electronic structure of bulk UOTe with LDA+U=4 eV highlighting the $C_4$-symmetry protected Dirac crossing along $\Gamma-A$ high symmetry direction. The color and size of the marker at each point represent the orbital type and contribution strength, respectively. (b) The change in electronic structure with the change of Hubbard U=3.5 eV. The symmetry-protected Dirac crossing is lifted, and a continuous gap appears along the $\Gamma-A$  direction, transiting the system to a weak AFM topological insulating state. (c) The signature of the weak topological insulating phase is determined by the evolution of the Wannier charge center on $k_z = 0$ and $k_z = \pi$ plane. The odd number of crossings made by the wannier charge center bands to the red dashed reference line indicates the the non-trivial character. (d) The unit cell of the bi-layer AFM UOTe with an open boundary along the c-direction. (e) The electronic structure of bi-layer UOTe in AFM ground state using hybrid functional HSE06 shows a global gap in the spectrum throughout the BZ. The breaking of effective time reversal symmetry in the 2D limit allows this phase to become a Chern insulator. (f) The surface state spectrum of bi-layer AFM UOTe in the Chern insulating phase shows a single chiral surface state to cross the Fermi energy within the bulk gap. (g) The calculated anomalous Hall conductivity shows the quantized $\sigma_{xy}$ component around the Fermi energy.}

\label{ext figure 6}
\end{center}
\end{figure*}

\sectionmark{EDX and SEM}
\begin{figure*}[ht]
\begin{center}
\includegraphics[width=14cm]{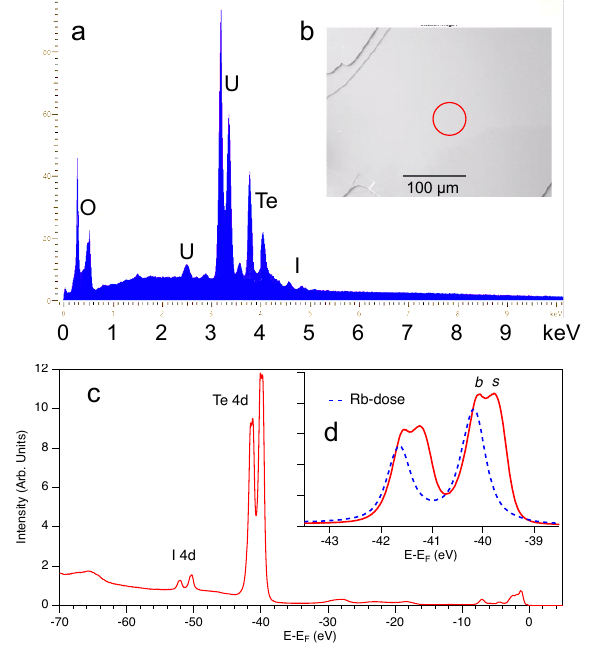}
\caption{
(a) SEM-EDX fluorescence spectrum showing the constituent O $K_\alpha$, U $M_\alpha$, Te $L_\alpha$ x-ray peaks as well as an impurity iodine  $L_\alpha$ component. 
(b) SEM image of the flat region being probed. The red circle represents the ARPES $\sim$50 $\mu$m beam spot size diameter.
(c) Wide-energy angle-integrated spectrum measured at 98 eV exhibiting a dominant Te 4d core-level and a weak Iodine 4$d$ impurity contribution. 
(d) The Te 4d core-level exhibits two bulk ($b$) and surface ($s$) components consistent with cleavage between Te-Te planes. The surface component is suppressed with surface deposition of alkali (Rb) atoms (dashed line).
}
\label{ext_edxsem}
\end{center}
\end{figure*}

\sectionmark{Table S1. EDX composition analysis }
\begin{table}[!h]
\centering
\caption{SEM-EDX composition analysis}
\begin{tabular}{ l cc c cc}
 \hline
 & Emission & Atomic & Deviation &  \\  
Element & Series & Percent & from UOTe  \\  
\hline
     U & M & 33.1\% & -    \\  
     O & K & 36.5\% & +3\%   \\  
     Te & L  & 28.4\% & -5\% & vacancy  \\   
     I & L & 2.0\% & +2\% & impurity \\   
\hline
\end{tabular}
\label{ext table 1}
\end{table}


\sectionmark{Linear response calculation of Hubbard U}
\begin{figure*}[ht]
\begin{center}
\includegraphics[width=12cm]{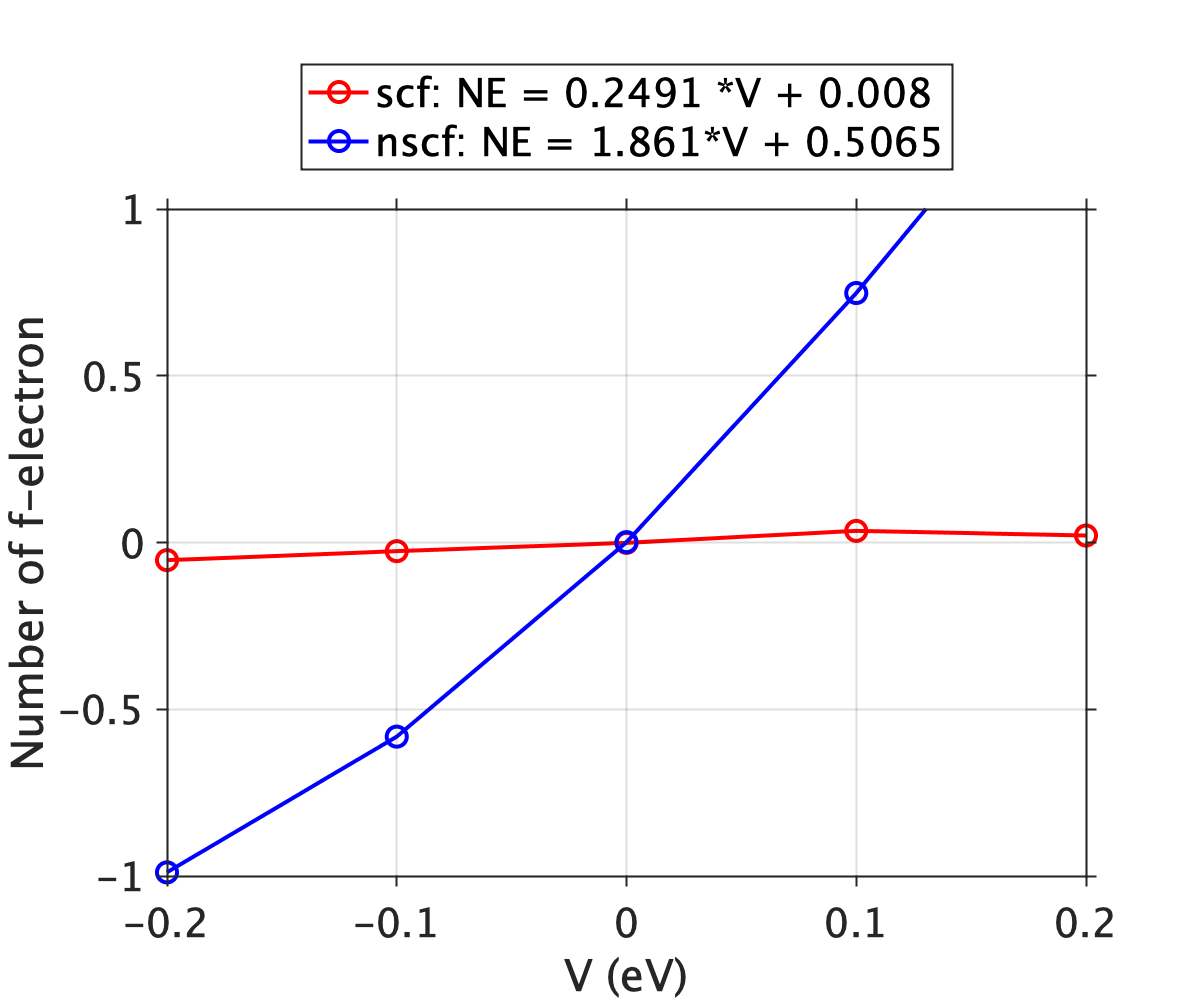}
\caption{
To estimate the value of Hubbard U of U-5f in UOTe we followed the linear response approach by Cococcioni et. al\cite{cococcioni_URPA}. The Hubbard U in this approach can be obtained by the following expression, $ U=\chi^{-1}-\chi_0^{-1} \approx\left(\frac{\partial N_i^{\mathrm{SCF}}}{\partial V_i}\right)^{-1}-\left(\frac{\partial N_i^{\mathrm{NSCF}}}{\partial V_i}\right)^{-1},$
where $\chi^{-1}$ and $\chi_0^{-1}$ are the self-consistent and non-self-consistent density response function due to the change of local potential at a particular site i. For UOTe the variation of Uranium f-electrons is shown in this figure, where the slope of the red(blue) curve is the measure of the self-consistent (non-self-consistent) response function. The calculated values of this response function for the self-consistent and non-self-consistent calculation are found to be 0.2491 $(eV)^{-1}$ and 1.861 $(eV)^{-1}$ respectively, which gives the effective interaction $U$ = 3.47~$eV$.
 }
\label{ext_u_rpa}
\end{center}
\end{figure*}

\sectionmark{Theory band folding}
\begin{figure*}[ht]
\begin{center}
\includegraphics[width=15cm]{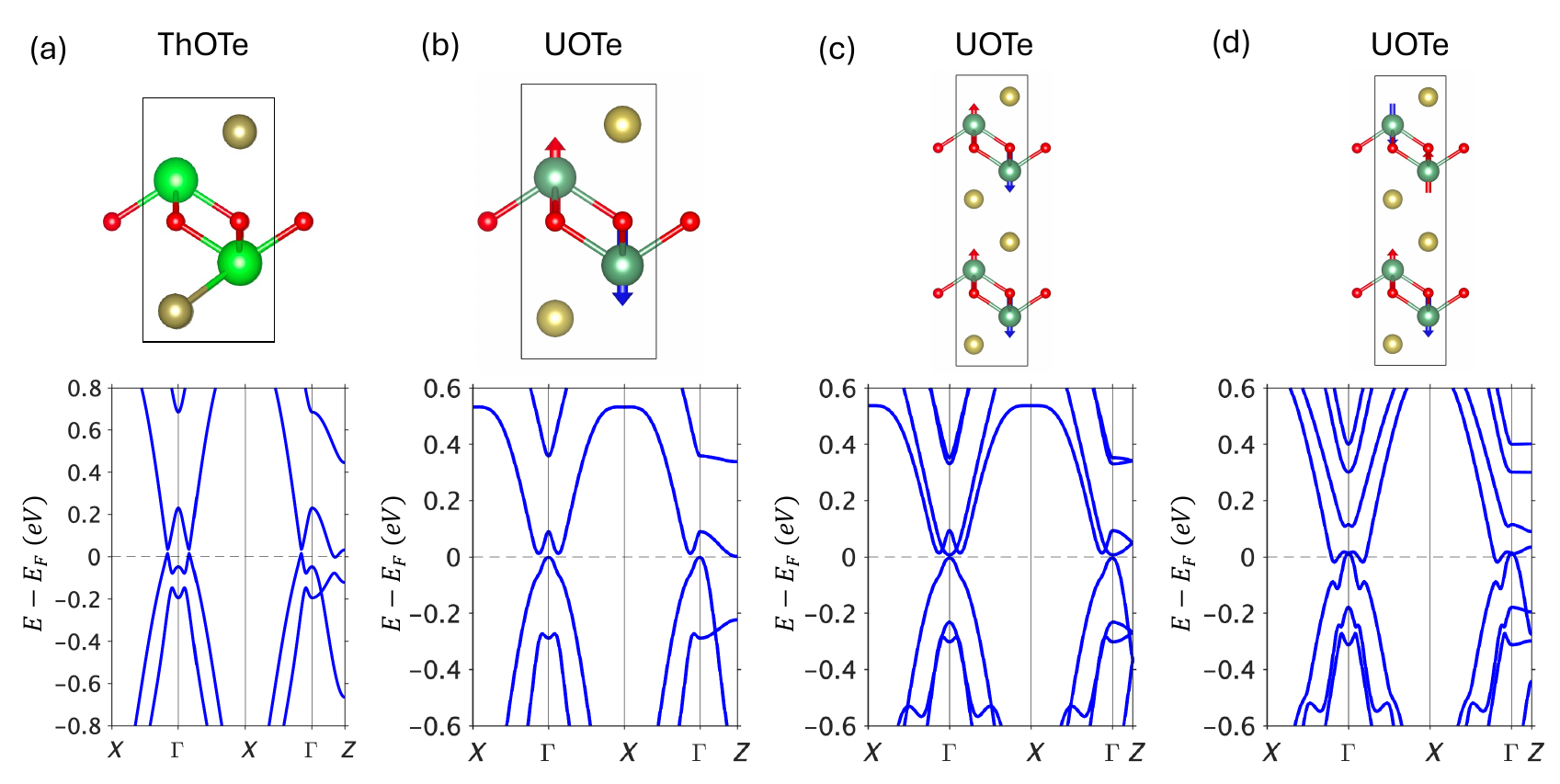}
\caption{
To illustrate the band folding picture of UOTe, we carried out a series of DFT calculation illustrating band evolution in different structural configurations to explain the band folding picture we presented in the main text. (a)-(f) Band structure calculation in ThOTe, UOTe $\uparrow \downarrow$ structure, UOTe doubled $\uparrow \downarrow \uparrow \downarrow$ structure, and UOTe $\downarrow \uparrow \uparrow \downarrow$ real magnetic ground state. Here (a) ThOTe represents the non-magnetic counterparts of UOTe, simulating the UOTe band structure above the Neel temperature. This band structure resembles the band in (b) UOTe $\uparrow \downarrow$ structure. When simply doubling the unit cell in (c), the band folding effect give rise to a doubled number of bands with band crossing points. Such band crossing points are further lifted when considering the real $\downarrow \uparrow \uparrow \downarrow$ magnetic ground states of UOTe, giving rise to the final band structure. From this series of band calculation, we can keep track of the band evolution and understand the nature of conduction band doubling near the Fermi level below the Neel temperature is due to Z-plane band folding effect originated from unit cell doubling of its antiferromagnetic structure. }
\label{ext_theory_bf}
\end{center}
\end{figure*}

\begin{figure}
    \centering
    \includegraphics[width=16cm]{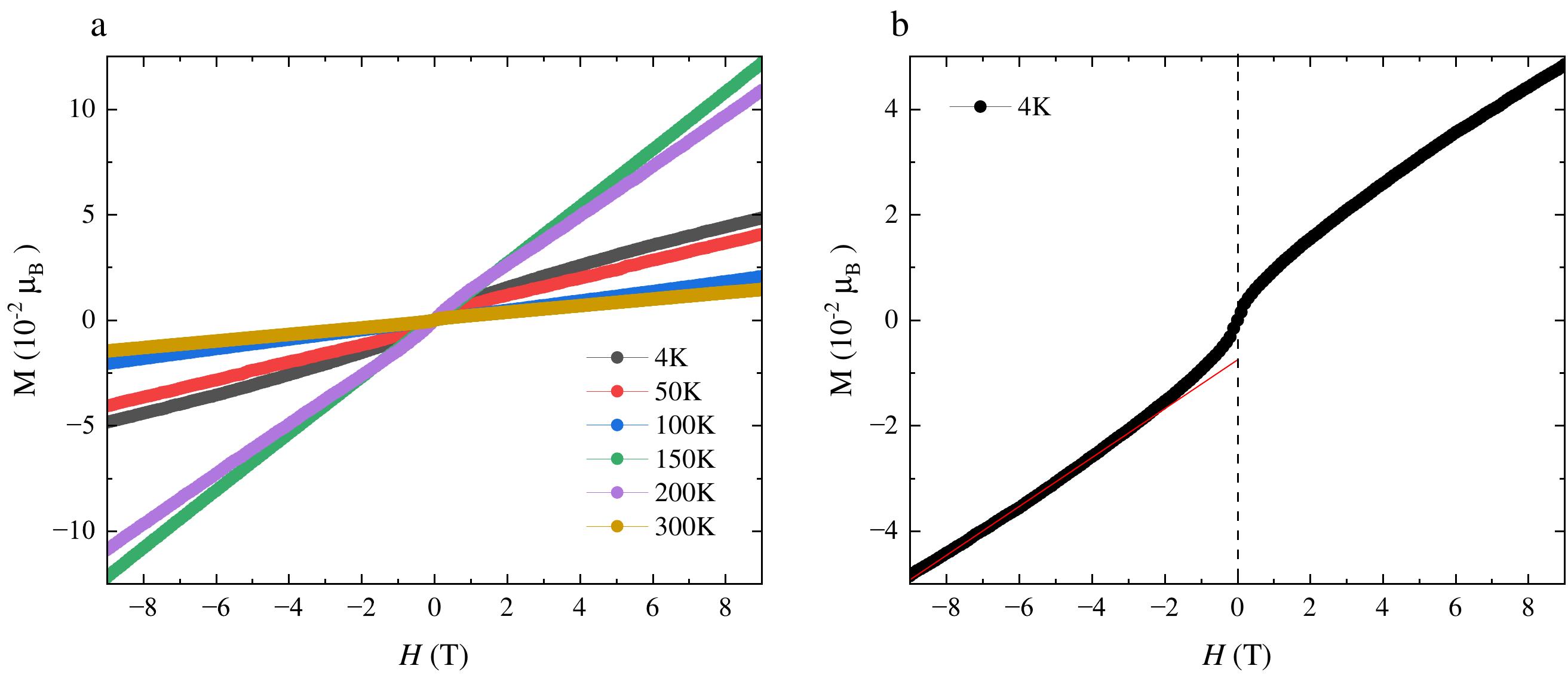}
    \caption{(a) Magnetization for UOTe shown for various temperatures between 4~K and 300~K. There is linear dependence and no saturation observed in magnetic fields up to 9~T above 100 K. A slight nonlinear dependence is observed in the low magnetic field is observed below 50 K, indicating ferromagnetic impurities. (b) The magnetization at 4~K, with linear extrapolation to zero field to show contribution from impurities. The ordered moment from ferromagnetic impurities is about 0.007 $\mu_B$}
    \label{ext_MvH}
\end{figure}
\end{document}